\documentclass[epj]{svjour}

\usepackage{graphicx}
\usepackage{fancyhdr}
\def\makeheadbox{{}}
\def\mytitle{My title} 
\def\myauthors{My name}
\def\mytype{My type of session}
\def\mysession{My session}
\def\mytitle{Searches for Squarks and Gluions with D0 Detector} 
\def\myauthors{Mansoora Shamim} 
\def\mytype{Contributed Talk} 
\def\mysession{Colliders - SUSY Phenomenology}
\begin{document}
\title{Searches for Squarks and Gluions with D0 Detector}
\author{Mansoora Shamim
\thanks{\emph{Email:shamimm@phys.ksu.edu}}
 ~for the D0 Collaboration}   
\institute{Kansas State University\\\\}
\date{}
\abstract{
Searches for supersymmetry in the framework of R-parity conserving models have been performed in data collected by the D0 detector at the Tevatron $p\bar{p}$ collider at a center-of-mass energy of 1.96 TeV. Topologies analyzed consist of acoplanar-jets, multijets, and leptons with large missing transverse energy in the final state. The 1 fb$^{-1}$ data shows good agreement with the standard model expectations.  The improved mass limits at 95$\%$ CL have been derived. In the first search, generic squarks and gluinos productions are investigated and lower limits of 375 GeV and 289 GeV are derived on the squarks and gluino masses, respectively, with $\tan$($\beta$) = 3, $A_0$ = 0 and $\mu <$ 0. In second analysis, squarks are searched for in a final state that has $\tau$ leptons accompanied by jets and missing transverse energy. This channel is explored for the first time at the Tevatron and excludes squark masses $<$ 366 GeV.    
Supergravity inspired models suggest the existence of light supersymmetric partners of the third generation quarks: a light stop for moderate values of $\tan(\beta)$, a light sbottom for large $\tan(\beta)$. If stop is the next-to-lightest supersymmetric particle, the expected decay channel is $\tilde{t} \rightarrow c \tilde{\chi}^0_1$ with $\tilde{\chi}^0_1$ assumed to be the lightest supersymmetric particle.   Using this channel, stop masses $<$ 149 GeV and neutralino masses $<$ 63 GeV are excluded.  If there is a mass hierarchy between stop, chargino and sneutrino masses, stop is expected to decay via $\tilde{t} \rightarrow b l \tilde{\nu}$ where $l$ could be an electron or muon.  The search performed with both electrons, muons and dimuon final states excludes stop mass $<$ 186 GeV for sneutrino mass of 71 GeV.
\PACS{
      {PACS-key}{describing text of that key}   \and
      {PACS-key}{describing text of that key}
     } 
} 
\maketitle
\section{Introduction}
\label{intro}
Supersymmetry (SUSY) may provide a solution to the fine tuning
problem if the supersymmetric particles have masses less than $1$ TeV,
strongly motivating the searches for SUSY objects at the Tevatron. Squarks and gluinos being the superpartners of strongly interacting particles in standard model should be copiously produced at Tevatron.  If squarks are lighter than gluinos they will tend to decay via $\tilde{q}\rightarrow q \tilde{\chi}_1^0$ and their pair production will yield an acoplanar dijet topology with missing transverse energy carried away by the neutralinos that are assumed to be lightest supersymmetric particles.  On the other hand if gluinos are lighter than squarks, their decay via $\tilde{g} \rightarrow q\bar{q} \tilde{\chi}_1^0$ will result in a final state having large number of jets and missing transverse energy.  \\
For moderate values of $\tan \beta$, SUSY models favor the existence of light supersymmetric partners of third generation quarks and lepton in particular $\tilde{\tau}$ and $\tilde{t}$.  Their decay will lead to lepton, jets and missing transverse energy in the final state.  The following sections describe several searches for squarks and gluinos performed by D0 using 1 fb$^{-1}$ of data.
\section{Generic Search for Squarks and Gluinos}
\label{sec:1}
\begin{figure}[htp]
\begin{center}
\includegraphics[width=0.35\textwidth,height=0.30\textwidth,angle=0]{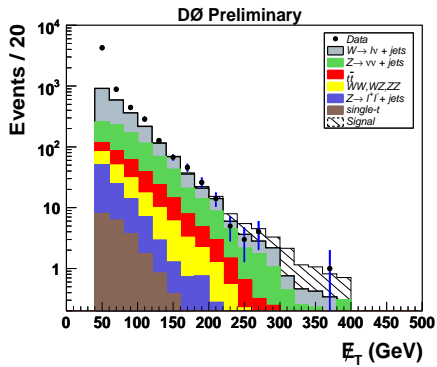}
\caption{$\not\!\!{E}_T$ distribution before applying the optimized cut on $H_T$ and $\not\!\!{E}_T$ in ``dijet analysis''.  }
\label{fig1}
\end{center}
\end{figure}

\begin{figure}[hbp]
\begin{center}
\centering
\includegraphics[width=0.45\textwidth,height=0.30\textwidth,angle=0]{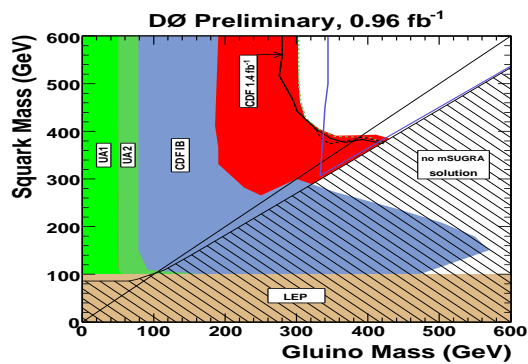}
\caption{In the squark gluino mass plane, regions excluded by the generic squark gluino analysis search.}
\label{fig2}
\end{center}
\end{figure}

This analysis searches for squarks and gluinos in dijet, three jets and four or more jets plus missing transverse energy final states without identifying the flavor of the jet\cite{ref1}.  The dijet analysis corresponds to the case where $m_{\tilde{q}} < m_{\tilde{g}}$ and their decay leads to at least two jets in the final state.  The three jet analysis refers to the situation when $ m_{\tilde{g}} \sim m_{\tilde{q}}$ while the presence of four or more jets describes the situation where $ m_{\tilde{g}} < m_{\tilde{q}}$ and is referred to as ``gluino analysis''.  The backgrounds for these channels are vector boson production in association with jets, multijet production, dibosons, single top, and top pair production.  
The selection criteria include cleaning up the events, selecting high $p_T$ jets, requiring large missing transverse energy and making sure that the jets are well separated not only from each other but also from the $\not\!\!{E}_T$ direction.\\
Figure \ref{fig1} shows the distributions for $\not\!\!\!{E}_T$ for ``dijet'' analysis after applying all analysis cuts except the cut on $~ \not\!\!\!{E}_T $.  After reducing the multijet production, signal to background ratio is enhanced by optimizing the cuts on two variables $\not\!\!{E}_T$ and $H_T$ where $H_T$ is scalar sum of $p_T$ of all jets.  The values of optimized cuts, final number of expected and observed events in three analyses are shown in Table \ref{tab:1}.  Since no excess is observed in the analyzed data over the expected SM background the lower limits on squark gluino masses are set.  Figure \ref{fig2} shows the region in squark and gluino mass plane excluded by this analysis. 
\begin{table}[htp]
\caption{Different analyses and corresponding number of expected background and observed events in generic squark gluino analysis search.}
\label{tab:1}      
\begin{tabular}{lcccc}
\hline\noalign{\smallskip}
Analysis & $H_T$ &$\not\!\!{E}_T$  &Background & Observed\\
&GeV &GeV &Expected& Events\\
\noalign{\smallskip}\hline\noalign{\smallskip}

``dijet'' &$>$ 300 & $>$ 225& 7.47 $\pm$ 1.06 & 5 \\
``3 jets'' & $>$ 400 & $>$ 150& 6.10 $\pm$ 0.37 & 6 \\
``gluino''&$>$ 300 & $>$ 100 &33.35 $\pm$ 0.81 & 24 \\
\noalign{\smallskip}\hline
\end{tabular}
\end{table}
\section{Search for Squarks with Taus in Final State}
\label{sec:2}
\begin{figure}[htp]
\begin{center}
\includegraphics[width=0.45\textwidth,height=0.20\textwidth,angle=0]{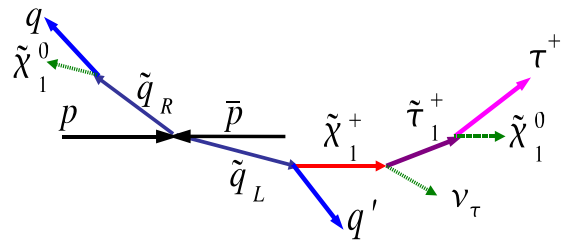}
\caption{Schematic diagram of pair production of squarks decaying into tau via exchange of chargino.}
\label{fig3}
\end{center}
\end{figure}

\begin{figure}[htp]
\begin{center}
\includegraphics[width=0.35\textwidth,height=0.30\textwidth,angle=0]{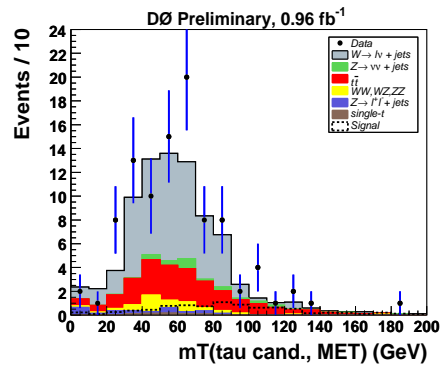}
\caption{Invariant mass reconstructed from a tau lepton and $\not\!\!E_{T}$.}
\label{fig4}
\end{center}
\end{figure}

\begin{figure}[hbp]
\begin{center}
\includegraphics[width=0.45\textwidth,height=0.30\textwidth,angle=0]{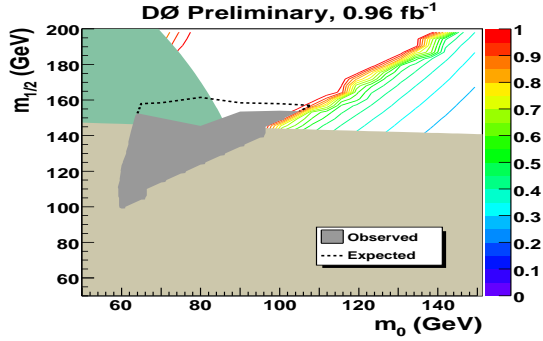}
\caption{Expected and observed limits in the $(m_o,m_{1/2})$ plane of the mSUGRA.}
\label{fig5}
\end{center}
\end{figure}

This analysis searches for squarks in multijet events accompanied by large missing transverse energy and at least one tau lepton decaying hadronically\cite{ref2}.  The search is motivated by the fact that the supersymmetric partner of $\tau$ lepton could be the lightest slepton due to the large mass splitting between the two scalar partners of $\tau$ leptons.  Figure \ref{fig3} shows the squark pair production and decay chain into tau final state.\\
Candidate events are selected by requiring high $p_T$ jets and a large amount of missing transverse energy which is well separated from jets.  The selection criteria also include requiring at least one hadronically decaying tau candidate identified via its 1-prong or 3-prong decay using a neural network algorithm optimized for tau identification.  The transverse momentum carried by $\tau$ lepton must exceed 15 GeV and $\tau$ should be well separated from the jets in the event.\\
The largest standard model background for this channel is top pair production.  The invariant mass reconstructed from a $\tau$ lepton and $\not\!\!{E}_T$ is shown in Figure \ref{fig4} which elaborates the fact that data is well described by the prediction from standard model processes.\\
The selection for two variables $H_T$ and $\not\!\!{E}_T$ is further optimized by minimizing the expected upper limit on the cross section in the absence of the signal.  Figure \ref{fig5} shows the resulting expected and observed lower limits in the ($m_{0},m_{1/2}$) plane of the mSUGRA framework with $\tan \beta$ = 15, $A_0$ = -2$m_0$ and $\mu < $ 0.  This analysis excludes squark masses (averaged over 10 squark species) up to 366 GeV at the highest value of $m_0$ of the limit contour.  
\section{Search for Supersymmetric Top Quark in Jets and $\not\!\!{E}_T$ Final State}
\label{sec:3}
\begin{figure}[hbp]
\begin{center} 
\includegraphics[width=0.25\textwidth,height=0.20\textwidth,angle=0]{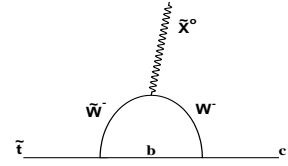}
\caption{Loop decay of stop into a charm quark and a neutralino.}
\label{fig6}
\end{center}
\end{figure}

\begin{figure}[htp]
\begin{center}
\includegraphics[width=0.35\textwidth,height=0.30\textwidth,angle=0]{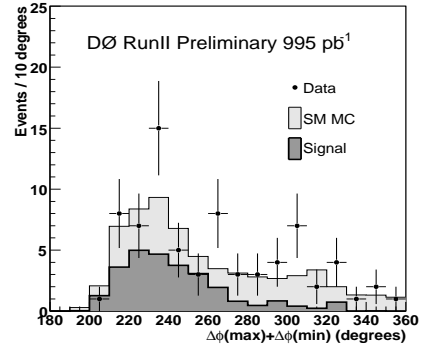}
\caption{The distribution of $\Delta \Phi_{max} +\Delta \Phi_{min}$ at the final stage of analysis. }
\label{fig8}
\end{center}
\end{figure}

\begin{figure}[hpb]
\begin{center}
\includegraphics[width=0.45\textwidth,height=0.33\textwidth,angle=0]{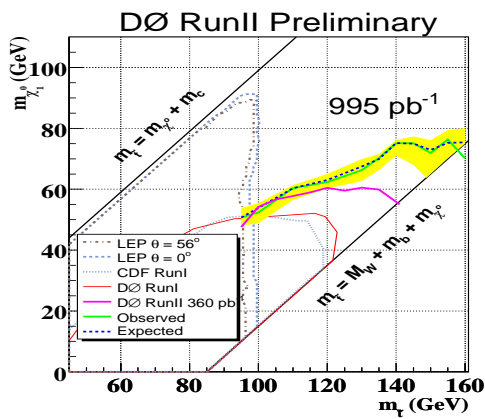}
\caption{Stop neutralino mass
plane excluded at 95 $\%$ confidence level by present search. The observed
(expected) exclusion contour is shown in green (blue) solid (dashed) line. The
yellow band represents the theoretical uncertainties on the production cross
section due to PDF and renormalization and factorization scale. Results from
previous searches are also shown. }
\label{fig9}
\end{center}
\end{figure}

The mass splitting between the supersymmetric partners of top quark being proportional to the mass of the top quark can make one of the stops lightest of all squarks.  In this search\cite{ref3}, stop is taken to be the next to the lightest supersymmetric particle and assumed to decay into a charm quark and a neutralino via loop decay shown in Figure \ref{fig6}.  The final state consists of two acoplanar charm jets and missing transverse energy.  The backgrounds to this channel are $W$ and $Z$ bosons produced in association with jets (both light and heavy flavor), multijet production, dibosons, single top and top pair production.\\
In order to calibrate $Z\rightarrow \nu\bar{\nu}$ + jets background, $Z\rightarrow e^+e^-$ + jets events are studied.  In order to correct for NNLO effects, $p_T$ of $Z$ boson is corrected using a reweighting function derived from the ratio of $Z$ boson $p_T$ in data to that in ALPGEN MC prediction. \\
The background prediction is estimated by normalizing the contribution from each channel to the observed number of $Z$ candidates in data.  This in turn reduces the 6.1$\%~\oplus$ 15$\%$ uncertainty on luminosity determination and theoretical cross section respectively to 4 $\%$ statistical uncertainty of $Z\rightarrow e^+e^-$ + $\geq$ 2 jets events.\\
After selecting the events on the basis of topological cuts, signal to background ratio is optimized by requiring at least one jet to be heavy flavor tagged using a dedicated neural network algorithm.  This algorithm makes use of information from non zero decay length and impact parameter for tracks coming from the decay of a charm hadron.  The final selection criteria are optimized using three variables $H_T, ~\not\!\!\!{E}_T $ and $\Delta \Phi_{max}$ +$\Delta \Phi_{min}$  where $\Delta \Phi_{max}$ and $\Delta \Phi_{min}$ are the maximum and minimum of the angle between jet and $ \not\!\!\!{E}_T $ direction respectively.  For each set of requirements, the expected value of signal confidence level $\langle CL_s \rangle$ under the hypothesis that only background was present was evaluated using all stop and neutralino mass combinations taking into account the systematic uncertainties.  For a given stop mass the set of requirements that excluded highest neutralino mass was selected.\\
Figure \ref{fig8} shows distribution for $\Delta \Phi_{max}$ +$\Delta \Phi_{min}$ variable at the final stage of the analysis.  The optimization criterion, expected background events and events observed in the data are shown in Table \ref{tab:2}.  The results of this analysis are shown in the final exclusion contour in stop neutralino mass plane in Figure \ref{fig9}.  Results from previous searches have also been plotted. 
\begin{table}
\caption{Optimized values of cuts, number of expected background and
observed data events.  A cut at $\not\!\!\!{E}_{T} > 70$ GeV was chosen in all cases.
The values of $H_{T}$ are in GeV while those for $\mathit{P}= \Delta \Phi_{max} +\Delta \Phi_{min}$ are in degrees.}
\label{tab:2}      
\begin{tabular}{lcccl}
\hline\noalign{\smallskip}
$m_{\tilde{t}}$& $H_{T} $&  $\mathit{P} $ & $\#$ & $\# $\\
& $>$ & $<$ &  observed & Expected\\
\noalign{\smallskip}\hline\noalign{\smallskip}
$95-130$ & $ 100$ & $ 260$ & $83 $ & $81.87\pm3.99^{+ 13.92}_{-14.06}$\\
$135-145 $ & $ 140$ &$ 300$ & $57 $ & $57.05\pm3.08^{+ 8.56}_{-8.64}$\\
$150-160$ &  $ 140$ &$ 320$ & $66 $ & $64.21\pm3.22^{+ 8.99}_{-9.08}$\\
\noalign{\smallskip}\hline
\end{tabular}
\end{table}
\section{Search for Supersymmetric Top Quark in Leptons, Jets and $\not\!\!{E}_T$ Final State}
\label{sec:4}
\begin{figure}
\centering
\includegraphics[width=0.30\textwidth,height=0.25\textwidth,angle=0]{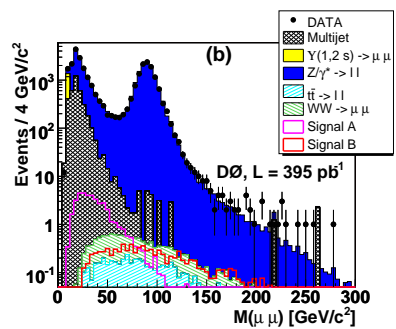}
\caption{Invariant mass of two most energetic muons.}
\label{fig10}
\end{figure}

\begin{figure}
\centering
\includegraphics[width=0.30\textwidth,height=0.25\textwidth,angle=0]{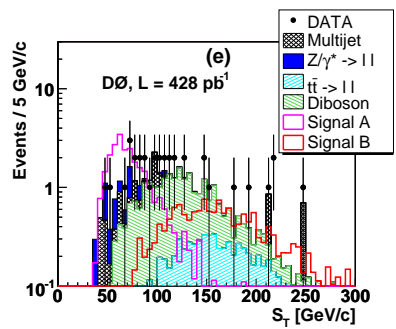}
\caption{Distribution of $S_T$ variable in electron muon channel.}
\label{fig11}
\end{figure}

\begin{figure}
\centering
\includegraphics[width=0.40\textwidth,height=0.35\textwidth,angle=0]{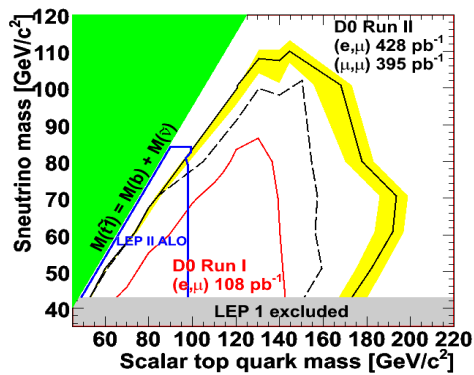}
\caption{95$\%$ C.L excluded region in the stop sneutrino mass plane for the observed (continous black line) and the average expected limit (discontinous black line); the yellow band shows the lower and upper bounds on the signal cross section variation.}
\label{fig12}
\end{figure}

If $\tilde{t}\rightarrow c \tilde{\chi}^0$ and $\tilde{t}\rightarrow b \tilde{\chi}^{\pm}$ modes are kinematically forbidden, then stop will decay into a lepton, $b$ quark and a sneutrino via $\tilde{t}\rightarrow b l \tilde{\nu}$.  The event signature will be two $b$ jets, a pair of isolated leptons and missing transverse energy.  A search is performed in two channels where in one channel both leptons are muons\cite{ref4} while in other channel one lepton is taken to be a muon and other an electron\cite{ref5}.  \\
The main background processes imitating the signal topology are $Z/\gamma^*, WW, t\bar{t}$ production and multijet background.  All but the later are estimated with MC.  The multijet background is estimated from data.  The selection criterion includes the requirement of high $p_T$ isolated leptons, missing transverse energy and jets.  The jets and missing transverse energy directions are required to be well separated from both leptons.  In the dimuon channel a cut on the invariant mass of two muons is placed in order to reduce $Z\rightarrow \mu\mu$ background.  The dimuon invariant mass is shown in Figure \ref{fig10}.\\
After making the basic selection, the dimuon analysis requires at least one jet to be heavy flavor tagged using the lifetime probability to be less than $1\%$.  The lifetime probability is the probability for the tracks in a jet to originate from the primary interaction point.  For heavy flavor jets this probability peaks towards zero while it is uniform for light flavor jets. At the last stage of the analysis both channels make use of the discriminating variables in order to separate signal from the background.  One of the discriminating variable ``$S_T$'' defined as the sum of the $p_T$ of electron, muon and $\not\!\!E_{T}$ is shown in Figure \ref{fig11}.  Instead of cutting on the discriminating variables the number of signal, background and data events are extracted which are then used to calulate the final limit on stop pair production cross section.  The final excluded region after combining two channels is shown in Figure \ref{fig12}.


\end{document}